\documentclass[twocolumn,aps,showpacs,preprintnumbers,amsmath,amssymb]{revtex4}

\usepackage{graphicx}
\usepackage{dcolumn}

\def\pos{p}
\def\clust{p}
\def\weightE{w}
\def\weightV{w}

\begin{document}

\title{Modularity clustering is force-directed layout}
\author{Andreas Noack}
\affiliation{Institute of Computer Science,
             Brandenburg University of Technology,
             03013 Cottbus, Germany
}

\date{\today}

\begin{abstract}
Two natural and widely used representations for the community structure
of networks are clusterings, which partition the vertex set
into disjoint subsets, and layouts, which assign the vertices
to positions in a metric space.
This paper unifies prominent characterizations
of layout quality and clustering quality,
by showing that energy models of pairwise attraction and repulsion
subsume Newman and Girvan's modularity measure.
Layouts with optimal energy are relaxations of, and are thus consistent with,
clusterings with optimal modularity,  which is of practical relevance
because both representations are complementary and often used together.
\end{abstract}

\pacs{89.75.Hc, 02.10.Ox}

\maketitle

\section{Introduction}

Many systems of scientific or practical interest are decomposable
into subsystems with strong internal
and relatively weak external interactions~\cite{Simon:1962};
for example, there are groups of friends or collaborators
in social networks, sets of topically related documents in hypertexts,
or blocs of interlocked countries in international trade.
If systems are modeled as networks, with the system elements as vertices
and their interactions as edges, then each subsystem corresponds to a so-called
{\em community}, a set of vertices with dense internal connections but sparse
connections to the remaining network.

Two widely used representations of networks are {\em layouts},
which assign the vertices to positions in a metric space,
and {\em clusterings}, which partition the vertex set into disjoint subsets.
Both representations can group densely connected vertices,
by placing them at nearby positions or in the same cluster,
and separate sparsely connected vertices,
by placing them at distant positions or in different clusters,
and can thus naturally reflect the community structure.
Requirements like the grouping of densely connected vertices
are often formalized as mathematical functions called {\em quality measures},
and the optimization of quality measures is a common strategy
for the computation of both layouts \cite{Brandes:1999a,DiBattistaEtAl:1999}
and clusterings 
\cite{Newman:EPJ2004,Gaertler:2005,Schaeffer:CSR2007,FortunatoCastellano:2007}.
Despite these commonalities, and although layouts and clusterings
are often used together as complementary representations of the same network,
there is no coherent understanding of layout quality and clustering quality.

This paper unifies Newman and Girvan's modularity~\cite{NewmanGirvan:2004},
a popular quality measure for clusterings,
with energy models of pairwise attraction and repulsion between vertices
(e.g., \cite{Brandes:1999a,DiBattistaEtAl:1999}),
a widely used class of quality measures for layouts.
After an introduction of the quality measures in Sec.~\ref{s:def},
Sec.~\ref{s:density} shows that layouts with optimal energy and clusterings
with optimal modularity represent the community structure similarly,
and Sec.~\ref{s:unification} demonstrates that modularity
actually {\em is} an energy model of pairwise attraction and repulsion,
if clusterings are considered as restricted layouts.
Section~\ref{s:appl} discusses the application of these results
for computing consistent clusterings and layouts.

\section{\label{s:def}Energy models and modularity}

Quality measures for representations of networks formalize what is considered
as a {\em good} representation, and allow to compute good representations
automatically using optimization algorithms.
Mathematically, a quality measure maps network representations to real numbers,
such that larger (or smaller) numbers are assigned to better representations,
and the best representations correspond to maxima (or minima) of the measure.
This section introduces two widely used quality measures,
namely energy models based on pairwise attraction and repulsion for layouts,
and Newman and Girvan's modularity measure for clusterings.

To obtain uniform and general formulations,
both measures are defined for {\em weighted} networks.
In a weighted network, each vertex~$v$ has a nonnegative real
{\em vertex weight}~$\weightV_v$, and each unordered vertex pair $\{u,v\}$
(including $u \mathop{=} v$) has a nonnegative real
{\em edge weight}~$\weightE_{\{u,v\}}$.
Intuitively, a vertex (or edge) of weight~$k$ can be thought of as a chunk
of $k$~vertices (or edges) of weight~$1$.
The commonly studied {\em un}weighted networks correspond
to the special case where the edge weights are either~$0$ (no edge) or~$1$,
and the vertex weights are~$1$.

\subsection{\label{ss:ldef}The $(a,r)$-energy model for layouts}

A {\em $d$-dimensional layout}~$\pos$ of a network maps each vertex~$v$
to a position~$\pos_v$ in~$\mathbb{R}^d$; 
it thereby assigns a {\em distance} to each vertex pair $\{u,v\}$, 
namely the Euclidean distance $\|\pos_u\mathop{-}\pos_v\|$ 
between the respective vertex positions.
So-called {\em energy models} are an important class of quality measures
for layouts.
In general, {\em smaller} energy indicates better layouts.
Because force is the negative gradient of energy,
energy models can also be represented as force systems,
and energy minima correspond to force equilibria.
For introductions to energy-based or force-directed layout,
see Refs.~\cite{Brandes:1999a,DiBattistaEtAl:1999}.

The most popular energy models for general undirected networks
are either similar to stress functions of multidimensional
scaling~\cite{BorgGroenen:1997},
or represent force systems of pairwise attraction and repulsion
between vertices.
Models of the former type (e.g., \cite{KamadaKawai:1989})
enforce that the distance of each vertex pair in the layout
approximates some prespecified distance, most commonly
the length of the shortest edge path between the vertices.
They will not be further discussed, because their layouts reflect
these path lengths rather than the community structure.

In models of the latter type,
adjacent vertices attract, which tends to group densely connected vertices,
and all pairs of vertices repulse,
which tends to separate sparsely connected vertices.
The strengths of the forces are often chosen to be proportional
to some power of the distance.
Formally, for a layout~$\pos$ and two vertices $u,v$ with $u \mathop{\ne} v$,
the attractive force exerted on~$u$ by~$v$ is
\[
\weightE_{\{u,v\}}   \: \|\pos_u\!\!-\!\pos_v\|^a \:
  \overrightarrow{\pos_u\pos_v}~,
\]
and the repulsive force exerted on~$u$ by~$v$ is
\[
\weightV_u\weightV_v \: \|\pos_u\!\!-\!\pos_v\|^r \:
  \overrightarrow{\pos_v\pos_u}~,
\]
where $\|\pos_u\mathop{-}\pos_v\|$ is the distance between $u$ and~$v$, $
\overrightarrow{\pos_u\pos_v}$
is the unit-length vector pointing from $u$ to~$v$,
and $a$ and~$r$ are real constants with $a \mathop{>} r$.

The condition $a\!>\!r$ ensures that the attractive force
between connected vertices grows faster than the repulsive force,
and thus prevents infinite distances except between unconnected components.
For most practical force models holds $a \mathop{\ge} 0$ and $r \mathop{\le} 0$,
i.e., the attractive force is non-decreasing and the repulsive force
is non-increasing with growing distance.
In the widely used force model
of Fruchterman and Reingold~\cite{FruchtermanReingold:1991},
$a \mathop{=} 2$ and $r \mathop{=} -1$.

By exploiting that force is the negative gradient of energy,
the force model can be transformed into an energy model, such that
force equilibria correspond to (local) energy minima.
For a layout~$\pos$ and constants $a,r \in \mathbb{R}$ with $a \mathop{>} r$,
the {\em $(a,r)$-energy} is
\begin{equation}
\sum_{\{u,v\}:\, u \ne v} \!\! \left(
    \weightE_{\{u,v\}}    \frac{\|\pos_u\!\!-\!\pos_v\|^{a+1}}{a+1}
~-~ \weightV_u \weightV_v \frac{\|\pos_u\!\!-\!\pos_v\|^{r+1}}{r+1}
\right)\!,
\end{equation}
where $\frac{\|\pos_u\!-\pos_v\|^{-1+1}}{-1+1}$ must be read as
$\ln \|\pos_u\!\!-\!\pos_v\|$ (because $x^{-1}$ is the derivative of $\ln x$).
The $(1,-3)$-energy model has been proposed by
Davidson and Harel \cite{DavidsonHarel:1996},
and the $(0,-1)$-energy model is known
as LinLog model~\cite{Noack:GD2003,Noack:JGAA2007}.

\subsection{\label{ss:cdef}The modularity measure for clusterings}

A {\em clustering}~$\clust$ of a network partitions the vertex set
into disjoint subsets called {\em clusters}, and thereby maps 
each vertex~$v$ to a cluster $\clust_v$.
Proposals of quality measures for clusterings are numerous and scattered
over the literature of diverse research fields; 
surveys, though non-exhaustive, are provided by 
Refs.~\cite{Gaertler:2005,Noack:JGAA2007,Schaeffer:CSR2007,Noack:Diss2007en}.

One of the most widely used quality measures was introduced 
by Newman and Girvan, and is called modularity.
It was originally defined for the special case
where the edge weights are either $0$ or~$1$ and the weight of each vertex is
its degree~\cite{NewmanGirvan:2004}, and was later extended
to networks with arbitrary edge weights~\cite{Newman:2004}.
(The {\em degree} of a vertex is the total weight of its incident edges,
with the edge weight from the vertex to itself counted twice.)
Generalized to arbitrary vertex weights, the {\em modularity}
of a clustering~$\clust$ is
\begin{equation}
\sum_{c \in \clust(V)} \left(
    \frac{\weightE_{\{c,c\}}}{\weightE_{\{V,V\}}}
  - \frac{\frac{1}{2}\weightV_c^2}{\frac{1}{2}\weightV_V^2}
\right),
\end{equation}
where $V$ is the set of all vertices in the network,
and $\clust(V)$ is the set of clusters;
the weight functions are naturally extended to sets of vertices or edges:
$\weightE_{\{c,c\}}$ is the total edge weight within the cluster~$c$,
and $\weightV_c$ is the total weight of the vertices in~$c$.

Intuitively, the first term of the modularity measure
is the {\em actual} fraction of intra-cluster edge weight.
In itself, it is not a good measure of clustering quality, because it takes
the maximum value for the trivial clustering where one cluster contains
all vertices.
This is corrected by subtracting a second term, which specifies
the {\em expected} fraction of intra-cluster edge weight in a network
with uniform density.
Thus modularity takes positive values for clusterings
where the total edge weight within clusters
is larger than would be expected if the network had no community structure.

\subsection{\label{ss:algs}Optimization algorithms}

Finding a minimum-energy layout or a maximum-modularity clustering 
of a given network is computationally hard;
in particular, modularity maximization was recently
shown to be NP-complete~\cite{BrandesEtAl:TKDE2008}.
In practice, energy and modularity are
almost exclusively optimized with heuristic algorithms
that do not guarantee to find optimal or near-optimal solutions.

An extensive experimental comparison of energy minimization algorithms
for network layout was performed 
by Hachul and J{\"u}nger~\cite{HachulJunger:JGAA2007};
however, most of the examined algorithms make fairly restrictive assumptions
about the optimized energy model.
More general and reasonably efficient is the force calculation algorithm
by Barnes and Hut~\cite{BarnesHut:1986}, whose runtime is $O(m + n \log n)$
per iteration for a network with $m$~edges (with nonzero weight) 
and $n$ vertices (assuming that the number of dimensions is small
and the vertex distances are not extremely nonuniform).
The number of iterations required for convergence typically grows sublinearly 
with~$n$.

Clustering algorithms for networks are surveyed in Refs.\
\cite{Newman:EPJ2004,DanonEtAl:2005,Schaeffer:CSR2007,FortunatoCastellano:2007}.
A relatively fast yet very effective heuristic for modularity maximization
is agglomeration by iteratively merging clusters (starting from singletons),
combined with single-level~\cite{SchuetzCaflisch:PRE2008} or multi-level 
\cite{Rotta:DA2008} refinement by iteratively moving vertices;
an efficient implementation requires a runtime of $O(m \log^2 n)$ 
(assuming $O(\log n)$ hierarchy levels in agglomeration
and $O(\log n)$ iterations through all vertices per level in refinement).

\section{\label{s:density}Energy models and modularity reveal communities}

A set of vertices is called a {\em community} if the density within the set is
significantly larger than the density between the set and the remaining network.
The {\em density between} two disjoint sets of vertices $T$ and $U$
is intuitively the quotient of the actual edge weight
and the potential edge weight between $T$ and $U$;
formally, it is defined as $\frac{\weightE_{\{T,U\}}}{\weightV_T\weightV_U}$,
where $\weightV_U$ is the total weight of the vertices in~$U$,
and $\weightE_{\{T,U\}}$ is the total edge weight between $T$ and~$U$.
Similarly, the {\em density within} a vertex set $U$
is $\frac{\weightE_{\{U,U\}}}{\weightV_U^2/2}$.
(This generalizes standard definitions of density
from graph theory~\cite{Diestel:2000} to weighted networks with self-edges.)

Existing theoretical results, which will be summarized and extended
in this section, already show that the community structure of a network
is reflected in layouts with optimal $(a,r)$-energy (for certain values of $a$
and~$r$) and in clusterings with optimal modularity.
What has previously escaped notice is the striking analogy:
{\em The separation of communities in an optimal layout
is inversely proportional to (some power of) the density between them,
and the separation of communities in an optimal clustering
reflects whether the density between them is smaller than a certain threshold.}
As an important limitation, the result for layouts will be derived only for
two communities, and cannot be expected to hold precisely for more communities.
Therefore, the consistency of $(a,r)$-energy layouts and modularity clusterings
will be revisited in Sec.~\ref{s:appl}, after further evidence
has been presented in Sec.~\ref{s:unification}.

In what appears to be the only previous work that formally relates
energy-based layout to modularity clustering~\cite{Noack:JGAA2007},
we did not established similarities between optimal layouts
and optimal clusterings, but only noted that the modularity measure
is mathematically similar to the density
(called normalized cut in~\cite{Noack:JGAA2007}), as both normalize
the actual edge weight with a potential or expected edge weight.

\subsection{\label{ss:ldensity}Representation of community structure
in layouts with optimal $(a,r)$-energy}

This subsection discusses how the distances in a layout
with optimal $(a,r)$-energy can be interpreted
in terms of the community structure of the network,
and how this interpretation depends on the parameters $a$ and~$r$.

\begin{figure}[htb]
  \centering
  \includegraphics[width=39mm]{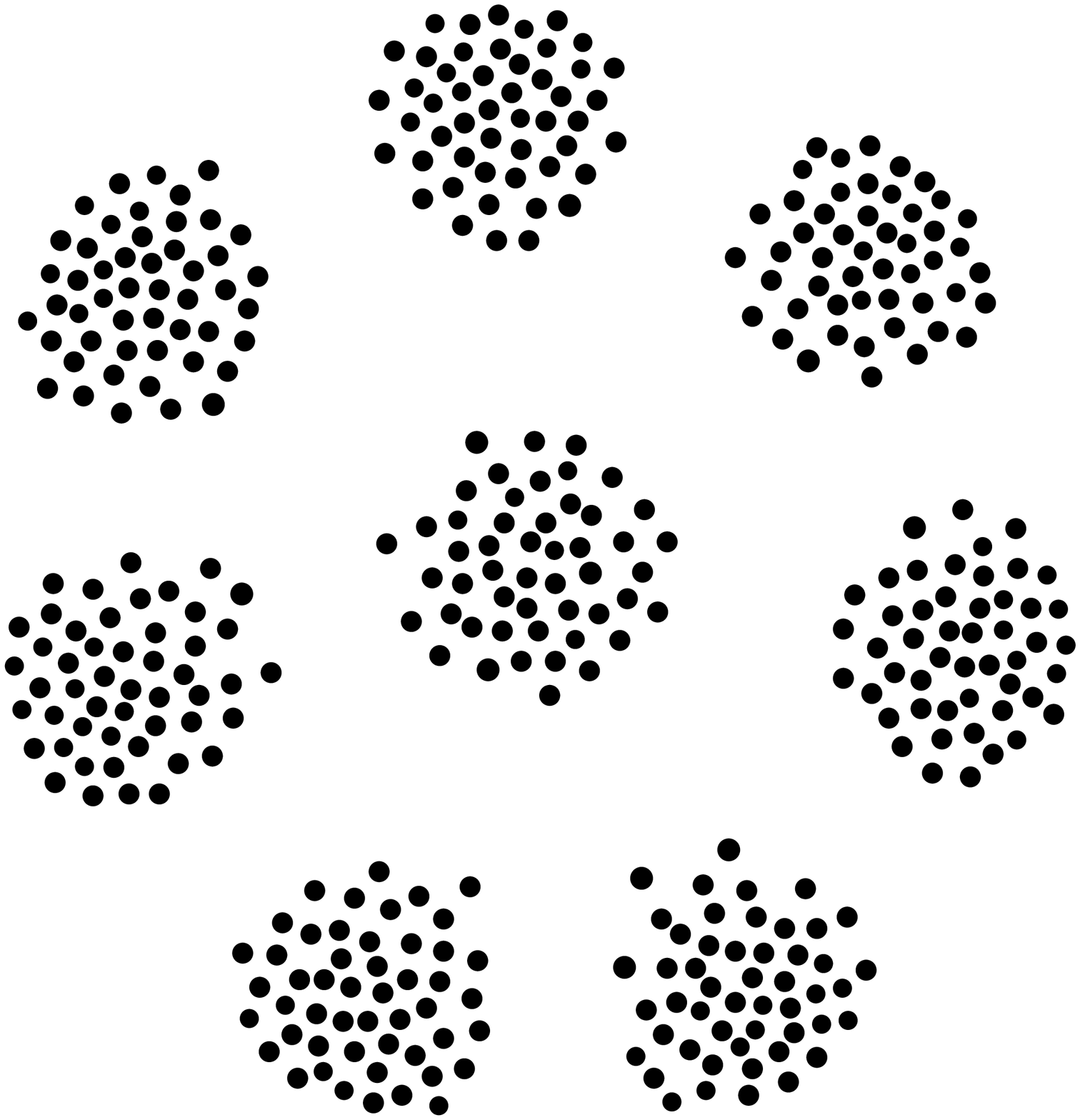} \hfill
  \includegraphics[width=39mm]{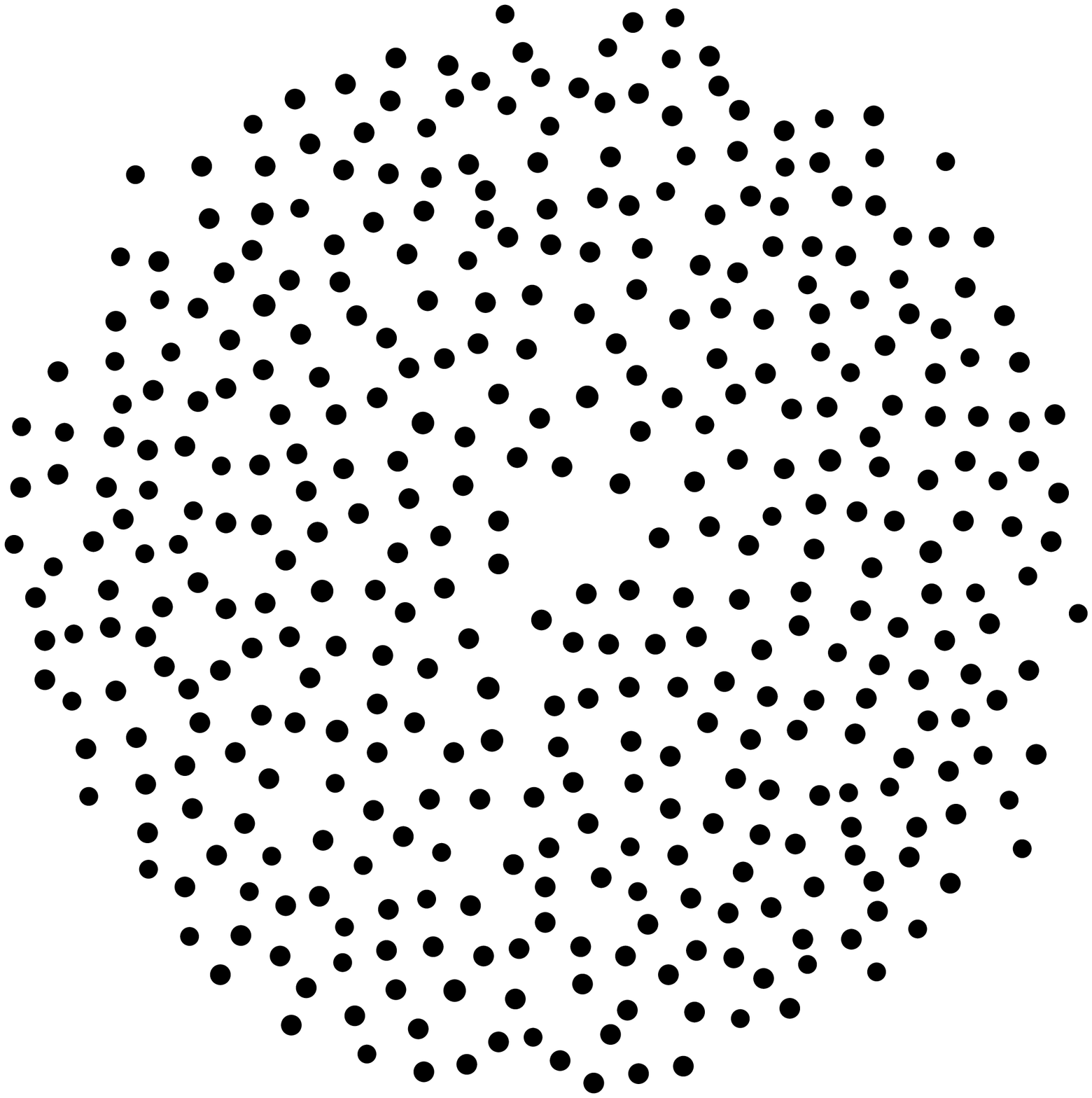}
  \caption{\label{f:random}%
    Layouts with small LinLog energy ($a \mathop{-} r \mathop{=} 1$)
    and with small Fruchterman-Reingold energy ($a \mathop{-} r \mathop{=} 3$)
    of a pseudo-random network with eight clusters
    (intra-cluster density~$1.0$, expected inter-cluster density~$0.2$).}
\end{figure}

For the simple case of a network with two vertices,
the minimum-energy layouts can be computed analytically
(Theorem~3 in \cite{Noack:JGAA2007}).
If the vertices $u$ and~$v$ have the distance~$d$, the $(a,r)$-energy is
\[
U(d)~~:=~~ \weightE_{\{u,v\}} \frac{d^{a+1}}{a+1}
      ~-~  \weightV_u \weightV_v \frac{d^{r+1}}{r+1}~.
\]
The derivative of this function is~$0$ at its minimum $d_0$, thus
\begin{eqnarray}
0 ~~=~~ U'(d_0) &=& \weightE_{\{u,v\}} d_0^a
                ~-~ \weightV_u \weightV_v d_0^r \nonumber \\
            d_0 &=& \left( \frac{\weightE_{\{u,v\}}}{\weightV_u\weightV_v}
                           \right)^{-\frac{1}{a-r}}~. \label{e:distdens}
\end{eqnarray}
Thus the distance of the two vertices in a layout with optimal $(a,r)$-energy
is the $-\frac{1}{a-r}$th power of the density between the vertices.
In particular, the distance is the inverse density
if $a \mathop{-} r \mathop{=} 1$, and the distance is almost independent
of the density if $a \mathop{-} r \gg 1$.
This impact of $a \mathop{-} r$ on the representation of the community 
structure is illustrated for a larger network in Fig.~\ref{f:random}.

Replacing the edge $\{u,v\}$ with two edges $\{u,t\}$ and $\{t,v\}$,
where $t$ is a new vertex with weight~$0$,
increases the optimal distance between $u$ and~$v$ by a factor of $2^{a/(a-r)}$.
Because the $(a,r)$-energy is only defined for $a \mathop{-} r \mathop{>} 0$,
the factor is~$1$ if $a \mathop{=} 0$, and greater than~$1$ if $a\mathop{>}0$.
This result has a significant implication,
given that the addition of~$t$ increases the path length between $u$ and~$v$
(from $1$ to~$2$ edges) without changing the density:
The optimal distance of $u$ and~$v$ depends only on the density,
and not on the path length, if $a \mathop{=} 0$
(as in the LinLog energy model), and increases with the path length
if $a\mathop{>}0$.

The results for networks with two or three vertices can be generalized,
at least as approximations, to larger networks.
In a network with clear communities, for example, the density within
the communities is (by definition) much greater than the density
between the communities,
and thus the intra-community distances in an optimal layout are much smaller
than the inter-community distances (unless $a \mathop{-} r$ is very large).
This can be approximated by assuming that the vertices of each community have
the same position, and thus by considering each community as one big vertex.
For networks with more than two communities, Eq.~(\ref{e:distdens})
cannot be expected to hold precisely for all pairs of communities,
because this would often imply distances that violate the triangle inequality.
Nevertheless, the qualitative reasoning generalizes:
Distances are less dependent on densities for large $a \mathop{-} r$,
and less dependent on path lengths for small~$a$.

\begin{figure}[htb]
  \begin{tabular}{c@{~~}|@{~~}c@{~~}|@{~~}c}
  & $a \mathop{=} 0$ & $a \mathop{=} 1$ \\[1mm]
  \hline
  \raisebox{5.5mm}{$r = 0$} &
  \raisebox{5.5mm}{violates $a \mathop{>} r$} &
  \parbox[b]{30mm}{\vspace{1mm}
                   \includegraphics[width=29mm]{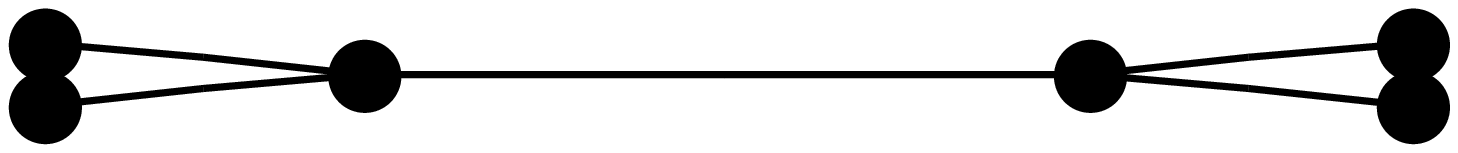}\\
                   \includegraphics[width=30mm]{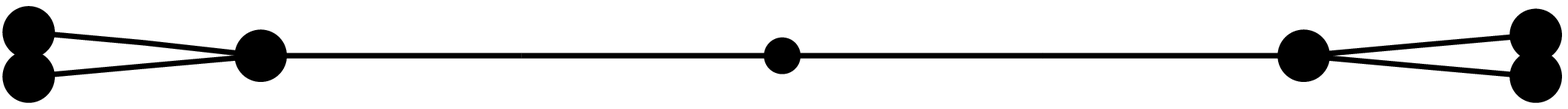}} \\[-0.5mm]
  \hline
  \raisebox{6mm}{$r \mathop{=} -1$} &
  \parbox[b]{30mm}{\vspace{2.5mm}
                   \includegraphics[width=30mm]{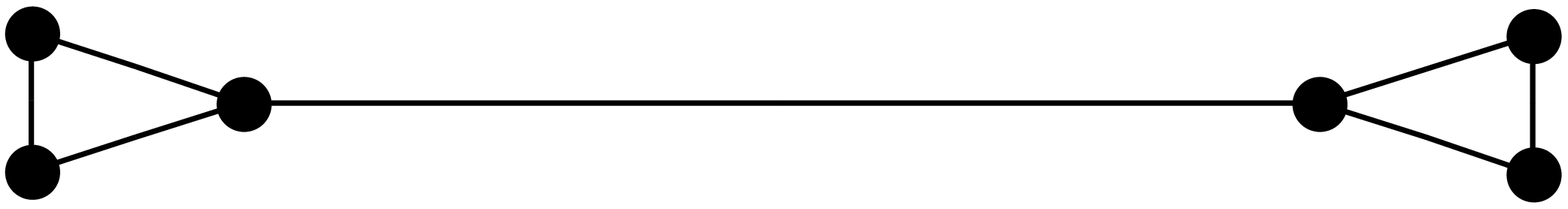}\\[2mm]
                   \includegraphics[width=30mm]{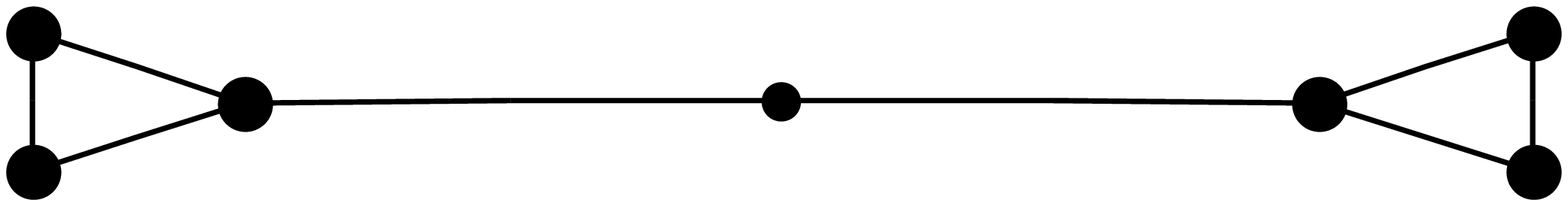}} &
  \parbox[b]{30mm}{\includegraphics[width=30mm]{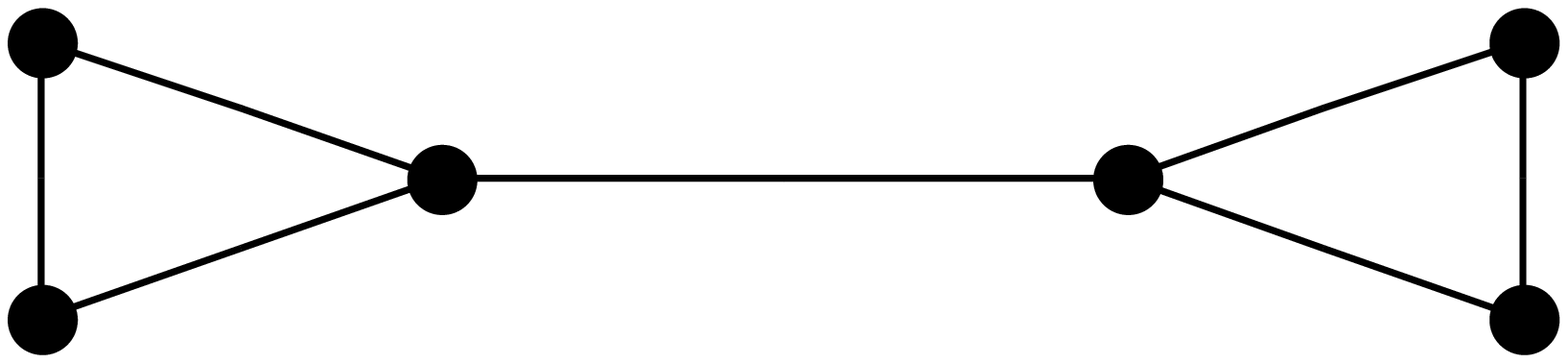}\\[2mm]
                   \includegraphics[width=30mm]{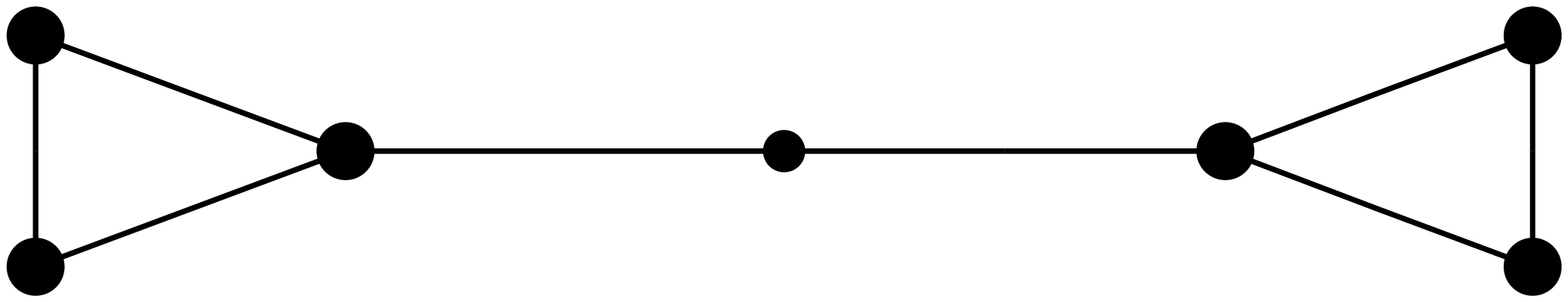}} \\[1mm]
  \end{tabular}
  \centering
  \caption{\label{f:paramexample}%
    Layouts with optimal $(a,r)$-energy for different values of $a$ and~$r$.
    All vertices and edges have weight~$1$, except for the small vertex
    between the triangles which has weight~$0$.}
\end{figure}

Figure~\ref{f:paramexample} illustrates the impact of the parameters
$a$ and~$r$ for two simple networks:
For $a \mathop{-} r \mathop{>} 1$ (bottom right), the two triangles
are less clearly separated than for $a \mathop{-} r \mathop{=} 1$
(bottom left and top right), and only for $a \mathop{=} 0$ (left),
the path length between the triangles does not affect their distance.

\begin{figure}[htb]
  \centering
  \includegraphics[width=58mm]{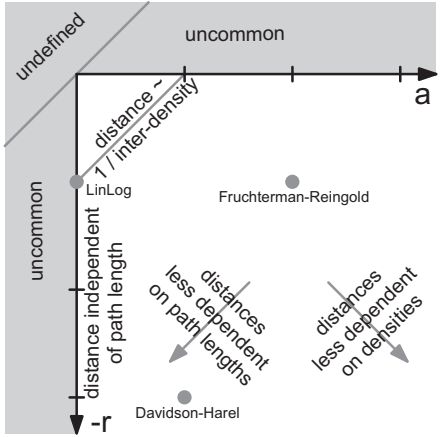}
  \caption{\label{f:paramsummary}%
    Impact of the parameters $a$ and~$r$
    on the optimal layouts of the $(a,r)$-energy model.}
\end{figure}

Figure~\ref{f:paramsummary} summarizes the results of this subsection.

\subsection{\label{ss:cdensity}Representation of community structure
in clusterings with optimal modularity}

Reichardt and Bornholdt~\cite{ReichardtBornholdt:PR2006}
observed that in a clustering with maximum modularity, the density
between any two clusters is at most the density within the entire network,
and the density between any two subclusters obtained by splitting
a cluster is at least the density within the network.
(Clusters may still have a smaller density than the network,
essentially because vertices without self-edges decrease the density
within their cluster but cannot be split.)
The argument is simple:
Joining two clusters $c$ and~$d$ with $c \mathop{\ne} d$
increases the modularity by
\[
    \frac{\weightE_{\{c,d\}}}{\weightE_{\{V,V\}}}
~-~ \frac{\weightV_c\weightV_d}{\frac{1}{2}\weightV_V^2}~,
\]
which is positive if and only if
\[
    \frac{\weightE_{\{c,d\}}}{\weightV_c\weightV_d}
~>~ \frac{\weightE_{\{V,V\}}}{\frac{1}{2}\weightV_V^2}~,
\]
i.e., if the density between $c$ and~$d$ is greater
than the density within the network.
In a clustering with maximum modularity, neither joining nor splitting clusters
may increase the modularity, which yields the claim.

These observations imply that the granularity of clusterings
with maximum modularity depends on the overall density within the network,
which may be undesirable for some applications.
For example, if the density within the network is sufficiently small,
then two dense subnetworks connected by only one light-weight edge
are joined into a single cluster, instead of forming two separate 
clusters~\cite{FortunatoBarthelemy:PNAS2007}.
Similarly, doubling a network (by adding a second copy of the same network)
halves its density, and thus generally coarsens the optimal clustering
instead of preserving it~\cite{BrandesEtAl:TKDE2008}.
Because such granularity-related issues are specific 
to discrete representations like clusterings,
they provide a major motivation for the supplementary 
(and sometimes even exclusive) use of continuous representations like layouts.

\section{\label{s:unification}Energy subsumes modularity}

{\em Modularity can be considered as a special case of $(a,r)$-energy.}
The first subsection formally derives this result,
and the second subsection explains how this derivation is facilitated
by the definitions of $(a,r)$-energy and modularity in Sec.~\ref{s:def},
which generalize previous definitions from the literature.

\subsection{Transformation of modularity into $(a,r)$-energy}

The modularity of a clustering~$\clust$ was defined
in Sec.~\ref{ss:cdef} as
\[
\sum_{c \in \clust(V)} \left(
    \frac{\weightE_{\{c,c\}}}{\weightE_{\{V,V\}}}
  - \frac{\frac{1}{2}\weightV_c^2}{\frac{1}{2}\weightV_V^2}
\right),
\]
i.e., as the difference of the actual fraction of intra-cluster edge weight
and the expected fraction of intra-cluster edge weight.

Because each edge is either intra-cluster or inter-cluster,
the fraction of intra-cluster edge weight
and the fraction of inter-cluster edge weight add up to~$1$:
\[
  \sum_{c \in \clust(V)}
    \frac{\weightE_{\{c,c\}}}{\weightE_{\{V,V\}}}
+ \sum_{\{c,d\} \subseteq \clust(V):\, c \ne d}
    \frac{\weightE_{\{c,d\}}}{\weightE_{\{V,V\}}}
~=~ 1\,;
\]
similarly, the corresponding expected fractions add up to~$1$.
Thus the modularity of~$\clust$ can be written
in terms of inter-cluster edge weights as
\begin{eqnarray*}
&&  \sum_{\{c,d\} \subseteq \clust(V):\, c \ne d} \left(
       - \frac{\weightE_{\{c,d\}}}{\weightE_{\{V,V\}}}
      + \frac{\weightV_c\weightV_d}{\frac{1}{2}\weightV_V^2}
    \right) \\
&=& - \sum_{\{u,v\} \subseteq V:\, \clust_u \ne \clust_v} \left(
        \frac{\weightE_{\{u,v\}}}{\weightE_{\{V,V\}}}
      - \frac{\weightV_u\weightV_v}{\frac{1}{2}\weightV_V^2}
    \right).
\end{eqnarray*}

Let $k$ be the number of clusters in~$\clust$.
Without changing the modularity of~$\clust$, the $k$~clusters
can be considered as positions in~$\mathbb{R}^{k-1}$,
such that each pair of different clusters has the distance~$1$.
(Intuitively, the $k$~clusters form the corners
of a regular $(k \mathop{-} 1)$-simplex with edge length~$1$;
a $(k \mathop{-} 1)$-simplex is the $(k \mathop{-} 1)$-dimensional analogue
of a triangle.)
Then the clustering~$\clust$ is a $(k\mathop{-}1)$-dimensional layout,
and the modularity of~$\clust$ can be rewritten as
\[
- \sum_{\{u,v\} \subseteq V:\, \clust_u \ne \clust_v} \! \left(
   \frac{\weightE_{\{u,v\}}}{\weightE_{\{V,V\}}}
     \|\clust_u\!\!-\!\clust_v\|
  -\frac{\weightV_u\weightV_v}{\frac{1}{2}\weightV_V^2}
    \|\clust_u\!\!-\!\clust_v\|
\right).
\]
The condition $\clust_u \mathop{\ne} \clust_v$ of the sum
can be dropped or replaced with $u \mathop{\ne} v$,
because it excludes only vertex pairs $\{u,v\}$
with $\|\clust_u \mathop{-} \clust_v\| \mathop{=} 0$.

Because the distances between the vertices are $0$ or~$1$,
the modularity of~$\clust$ equals
\[
- \sum_{\{u,v\}:\, u \ne v} \!\!\left(\!
    \frac{\weightE_{\{u,v\}}}{\weightE_{\{V,V\}}}
    \|\clust_u\!\!-\!\clust_v\|^{a+1}
  - \frac{\weightV_u\weightV_v}{\frac{1}{2}\weightV_V^2}
    \|\clust_u\!\!-\!\clust_v\|^{r+1}
\!\right)
\]
for all $a,r \in \mathbb{R}$ with $a \mathop{>} -1$ and $r \mathop{>} -1$.
This is the negative $(a,r)$-energy,
except for the constant factors in the attraction term and the repulsion term,
which change only the scaling of the optimal layouts.

\subsection{Prerequisites of the transformation}

The transformation of modularity into $(a,r)$-energy in the previous subsection
is based on the definitions of the measures in Sec.~\ref{s:def},
which generalize previous definitions from the literature in several respects.

First, the goal of most energy-based layout techniques
is to produce easily readable box-and-line visualizations,
which differs from and even conflicts with producing
faithful representations of the community structure.
The classic energy models of Eades~\cite{Eades:1984},
Fruchterman and Reingold~\cite{FruchtermanReingold:1991},
and Davidson and Harel~\cite{DavidsonHarel:1996} primarily reward
the conformance to aesthetic criteria like small edge lengths
and uniformly distributed vertices,
and thus often prevent the clear separation of sparsely connected vertices
and the clear grouping of densely connected vertices (see Fig.~\ref{f:random}).
The design and evaluation of energy models
with the explicit purpose of representing the community structure started
only recently with the LinLog model~\cite{Noack:GD2003,Noack:JGAA2007}.
Technically, the classic energy models are, or are similar to, instances
of the $(a,r)$-energy model where the difference $a\mathop{-}r$ is fixed
and too large; the $(a,r)$-energy model is parameterized with this difference.

Second, most existing energy models are designed
to strongly discourage the placement of several vertices on the same position,
while clusterings may place many vertices in the same cluster.
Technically, existing energy models are not mathematically equivalent
to modularity because the exponent of the distance in the repulsion energy
is fixed and too small;
the $(a,r)$-energy model is parameterized with this exponent.

Third, the modularity measure and most energy models were originally defined
for networks without vertex weights.
The vertices are implicitly weighted with~$1$ in most classic energy models
(e.g., \cite{Eades:1984,FruchtermanReingold:1991,DavidsonHarel:1996}), and
with their degree in the original modularity measure~\cite{NewmanGirvan:2004}.
It was only recently observed that degree-weighting may also improve
the readability and interpretability of energy-based layouts
\cite{Noack:GD2005,Noack:JGAA2007}.
The definitions of $(a,r)$-energy and modularity in Sec.~\ref{s:def}
are generalized to arbitrary vertex weights, and thus subsume
both degree weights and unit weights.

\subsection{\label{ss:urel}Related work}

In the analysis of dissimilarity matrices, the computation of clusterings
and layouts with identical quality measures is fairly common
(e.g., \cite{Shepard:1980,CarrollPruzansky:1980}).
The trick is to represent both clusterings and layouts
of dissimilarity matrices as dissimilarity matrices:
The dissimilarity of two objects in a layout can be defined
as their Euclidean distance (as for networks),
and the dissimilarity of two objects in a clustering can be defined
as the average dissimilarity of the objects in their clusters
(unlike for networks, which specify no dissimilarities for their vertices).
With this common representation of clusterings and layouts,
it is easy to design common quality measures.

For networks, there appear to be no previous proposals of using
identical quality measures for both clusterings and layouts.
Some clustering algorithms compute layouts as intermediate results,
for example eigenvector-based heuristics
for modularity clustering \cite{WhiteSmyth:2005long,Newman:PR2006}
and approximation algorithms for some related partitioning
problems \cite{LinialEtAl:1995,AumannRabani:1998,AroraEtAl:2004long},
but these layouts are not intended to be useful on their own.

\section{\label{s:appl}Optimal-energy layouts conform to 
  optimal-modularity clusterings}

Clusterings and layouts complement each other
as representations for the community structure of networks.
Layouts are limited to two or three dimensions in practice,
and thus cannot faithfully represent inherently high-dimensional structures,
but they may show crucial details that are missing in clusterings:
\begin{itemize}
\item the density between clusters, and more generally,
  the relationship between clusters,
  e.g., whether their separation is clear or fuzzy,
  and which vertices form their interface,
\item the density within clusters, and more generally,
  the internal structure of clusters,
  e.g., whether a dense cluster is composed of even denser subclusters,
\item the density between vertices and clusters,
  e.g., whether a vertex is central or peripheral to its cluster,
  or whether the assignment of a vertex to a cluster is rather arbitrary
  because it is closely related to several other clusters.
\end{itemize}

However, a layout only permits these interpretations
if it is {\em consistent} with the respective clustering, i.e., if the layout
and the clustering group the vertices according to the same criteria.
In previous works, some authors nonetheless consider vertex groups
in arbitrary force-directed layouts as clusters,
while others rightly note that they have no reasons to suppose
that such interpretations are valid.
Sections \ref{s:density} and~\ref{s:unification} finally provide such reasons,
as summarized in the following subsection.

\subsection{\label{ss:apre}Evidence}

Section~\ref{s:unification} showed that for clusterings with $k$~clusters,
considered as restricted $(k\mathop{-}1)$-dimensional layouts,
the $(a,r)$-energy model is equivalent to the modularity measure
if $a \mathop{>} -1$ and $r \mathop{>} -1$.
Thus (unrestricted) layouts with optimal $(a,r)$-energy are relaxations
of clusterings with optimal modularity
if (a) the layouts have at least $k\mathop{-}1$ dimensions,
and (b) $a \mathop{>} -1$ and $r \mathop{>} -1$.

Concerning condition~(a), the dimensionality of layouts can be somewhat reduced
without large changes of the pairwise vertex distances,
and thus without large changes of the $(a,r)$-energy.
Hence the consistency of optimal layouts
and optimal clusterings does not break down immediately if the layout
has less dimensions than the clustering has clusters.

Condition~(b) does not imply that layouts
with optimal $(a,r)$-energy closely resemble clusterings
with optimal modularity precisely for $a \mathop{>} -1$ and $r \mathop{>} -1$.
On the one hand, the condition $r \mathop{>} -1$ is necessary
for clusterings to permit the assignment of several vertices
to the same cluster, but not for layouts which may group vertices
without placing them on exactly the same position.
On the other hand, the precise values of $a$ and~$r$ hardly matter
for clusterings where the distance between vertices is either $0$ or~$1$,
but were shown to be important for layouts in Sec.~\ref{s:density}.
Considering the results of Sec.~\ref{s:density},
$(a,r)$-energy layouts most closely resemble modularity clusterings if
\begin{itemize}
\item $a \mathop{>} r$, $a \mathop{\ge} 0$, and $r \mathop{\le} 0$
  (by the definition of $(a,r)$-energy),
\item $a \mathop{\approx} 0$, such that distances
  do not reflect path lengths, and
\item $a \mathop{-} r \mathop{\approx} 1$,
  or at least $a \mathop{-} r \mathop{\not\gg} 1$,
  such that distances reflect densities.
\end{itemize}

\subsection{Examples}

The purpose of this subsection is to illustrate the consistency
of $(a,r)$-energy layouts and modularity clusterings, and the benefits
of this consistency, for several real-world networks.
It should be stressed that the purpose is {\em not}
to validate the $(a,r)$-energy model or the modularity measure,
which are already widely used and discussed in many previous works;
and the purpose is {\em not} to {\em prove} the consistency
of $(a,r)$-energy layouts and modularity clusterings,
because the mathematical evidence summarized in the previous subsection
is more general than any number of examples.

The example networks are listed in Table~\ref{t:anetworks}.
The weight of each vertex is set to its degree,
as in the original modularity measure~\cite{NewmanGirvan:2004}
and in the edge-repulsion LinLog energy model~\cite{Noack:JGAA2007}.
In visualizations, each vertex is represented as a box, its degree (weight)
as area of the box, and its cluster membership as shape of the box.

\begin{table}[th]
\centering
\caption{\label{t:anetworks}Example networks}
\begin{ruledtabular}
\begin{tabular}{lrl}
Name             & Size & Source \\
\hline
Karate Club          &  $34$  & \cite[Figure~3]{Zachary:1977};
                                unweighted version \\
  && used in \cite{NewmanGirvan:2004,Newman:PNAS2006,BrandesEtAl:TKDE2008}\\
Book Co-Purchase     & $105$  & V.~Krebs, provided M.~Newman\footnote{
         {\tt http://www-personal.umich.edu/\symbol{126}mejn/netdata/}};\\
  && also used in \cite{Newman:PNAS2006,BrandesEtAl:TKDE2008}\\
Food Classification  &  $45$  & \cite{RossMurphy:1999},
         published in \cite[Table~5.1]{HubertEtAl:2001}\\
World Trade          &  $66$  & World Bank\footnote{
         Trade and Production Database at {\tt http://www.worldbank.org}}\\
\end{tabular}
\end{ruledtabular}
\end{table}

As motivated in the previous subsection,
the parameters of the energy model are set to
$a \mathop{=} 0$ and $r \mathop{\in} \{-2,-1.5,-1\}$,
with $r \mathop{=} -2$ for networks with very nonuniform density
($\mbox{modularity} \mathop{>} 0.5$),
and $r \mathop{=} -1$ for networks with fairly uniform density
($\mbox{modularity} \mathop{<} 0.3$).
The variation of~$r$ improves the readability
by ensuring that vertices are not placed too closely,
but otherwise does not affect the grouping of the vertices.

Because the exact optimization of $(a,r)$-energy and modularity
is computationally hard, the presented layouts and clusterings
are not guaranteed to be optimal (except for the clustering
of the Book Co-Purchase network~\cite{BrandesEtAl:TKDE2008}),
but are the best known representations.
The Java program used for generating these representations
is freely available~\footnote{\tt http://code.google.com/p/linloglayout/}.
It employs the Barnes-Hut algorithm for energy minimization,
and agglomeration with multi-level refinement for modularity maximization
(see Sec.~\ref{ss:algs}).

\begin{figure}[htb]
  \centering
  \includegraphics[width=86mm]{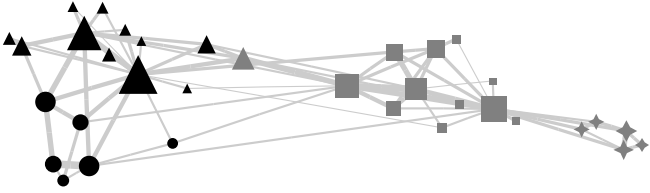}
  \caption{\label{f:karate}$(0,-1.5)$-energy layout and modularity clustering
    (represented by shapes) of the Karate Club network.
    The modularity of the clustering is~0.445.
    Gray boxes represent members who left the club after the instructor
    was fired.}
\end{figure}

In the Karate Club network (Fig.~\ref{f:karate}),
each vertex represents a member of a karate club,
and the edge weight of each vertex pair specifies
the number of contexts (like university classes, bars, or karate tournaments)
in which the two members interacted.
The main vertex groups in the $(0,-1.5)$-energy layout coincide with
the four clusters of the modularity clustering,
and the layout correctly indicates that joining triangles and circles
into a single cluster is almost as good as separating them
(modularity $0.435$ vs.~$0.445$).
The clustering and the layout both segregate the members
who left the club after the instructor was fired (gray boxes),
with the exception of one member who followed the instructor
mainly to preserve his chance for the black belt.

\begin{figure}[htb]
  \centering
  \includegraphics[width=86mm]{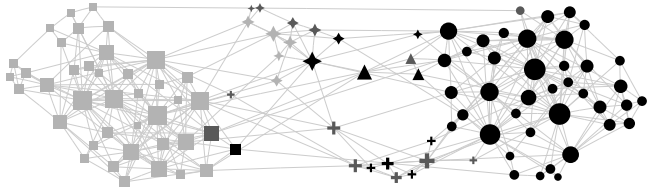}
  \caption{\label{f:polbools}$(0,-2)$-energy layout and modularity clustering
    of the Book Co-Purchase network.  The modularity is~0.527.
    Shades represent the classification
    as liberal (light gray), neutral (dark gray), or conservative (black).}
\end{figure}

In the Book Co-Purchase network (Fig.~\ref{f:polbools}),
the vertices represent books on US politics, and edges of weight~$1$
connect books that were frequently purchased together.
The clusters are generally well-separated in the layout;
a few members of the smaller central clusters are placed closely
to one of the two large clusters, which correctly indicates
that they are densely connected with parts of these large clusters,
and their assignment to a smaller cluster is a close decision.
The clustering and the layout, especially their two main groups,
conform well to Newman's classification~\cite{Newman:PNAS2006} of the books
as liberal (light gray), neutral (dark gray), or conservative (black);
the layout is more suitable to represent
the liberal-to-conservative ordering of the books.

\begin{figure}[htb]
  \centering
  \includegraphics[width=86mm]{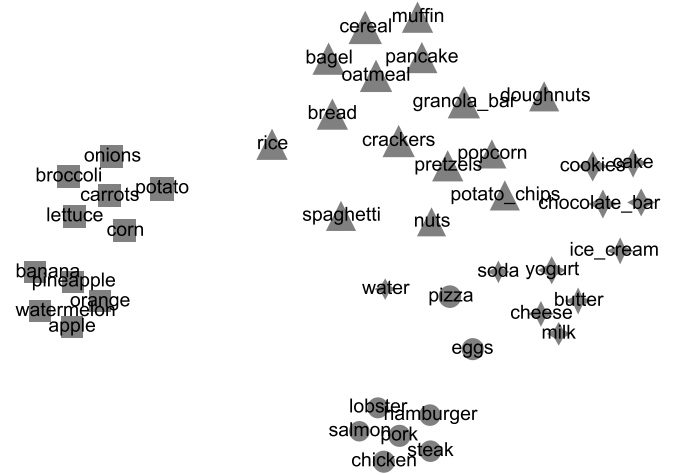}
  \caption{\label{f:food}$(0,-1.5)$-energy layout and modularity clustering
    of the Food Classification network.
    The modularity of the clustering is~0.402.
    (The edges are elided to avoid clutter.)}
\end{figure}

The Food Classification network (Fig.~\ref{f:food}) represents
the categorizations of 45~foods by 38~subjects of a psychological experiment, 
who were asked to sort the foods into as many categories 
as they wished based on perceived similarity.
Each vertex represents a food, and the edge weight of each vertex pair
is the number of subjects who assigned the corresponding foods
to the same category.
The clusters correspond well to groups in the layout,
but the layout also indicates that the borders between some clusters
are rather fuzzy (e.g., between snacks and sweets),
that some clusters could be split into subclusters 
(e.g., fruits and vegetables),
and that some foods cannot be clearly assigned to a single cluster
(e.g., water, spaghetti).
The grouping in both the clustering and the layout largely conforms
to common food categories.

\begin{figure}[htb]
  \centering
  \includegraphics[width=86mm]{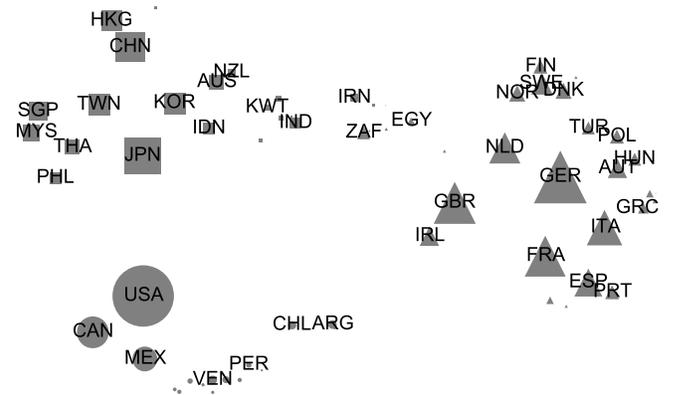}
  \caption{\label{f:trade}$(0,-1)$-energy layout and modularity clustering
    of the World Trade network.  The modularity of the clustering is~0.275.
    (The edges are elided to avoid clutter.)}
\end{figure}

The World Trade network (Fig.~\ref{f:trade}) models the trade
between 66~countries in the year~1999.
The vertices represent countries, and the edge weight of each vertex pair
specifies the trade volume between the corresponding countries in US~dollar.
The clustering and the layout both group the countries of the three major
economic areas (East Asia / Australia, America, and Europe).
The layout also reflects that countries like {\sf IRN} and~{\sf EGY} 
cannot be clearly assigned to either the East Asian or the European group,
and shows many smaller groups of closely interlocked countries 
like {\sf CHN} and~{\sf HKG}, {\sf AUS} and~{\sf NZL}, 
{\sf GBR} and~{\sf IRL}, and the Nordic countries.

\section{Conclusion}

As representations for the community structure of networks,
layouts subsume clusterings, thus quality measures
for layouts subsume quality measures for clusterings,
and in fact prominent existing quality measures for layouts --
namely, energy models based on the pairwise attraction and repulsion
of vertices -- subsume a prominent existing quality measure for clusterings --
namely, the modularity measure of Newman and Girvan.
This result has implications for the entire lifecycle of quality measures:
\begin{itemize}
\item Design: New and existing quality measures for layouts may be applied
  to clusterings and vice versa.  For example, recent extensions
  of the modularity measure to directed networks~\cite{LeichtNewman:PRL2008}
  and bipartite networks~\cite{Barber:PRE2007}
  can be directly generalized to energy models for layouts.
\item Evaluation:
  The evaluation of quality measures for clusterings and layouts
  can be partly unified, i.e., performed without distinguishing
  between clusterings and layouts.
  This has been demonstrated in~\cite{Noack:Diss2007en}
  with a computation of the expected measurement value for networks
  with uniform expected density,
  a particularly important analysis technique
  \cite{GaertlerEtAl:AAIM2007,Newman:PR2006,Noack:JGAA2007}.
\item Optimization:
  Components of clustering algorithms may be reused in layout algorithms
  and vice versa, for example the agglomeration (coarsening) phase 
  of multi-level heuristics.
  Moreover, energy-based layout algorithms might serve as initial stage
  of clustering algorithms, similarly to eigenvector-based layout algorithms
  in existing approaches (see Sec.~\ref{ss:urel}).
\item Application: Unified quality measures help to ensure the consistency
  of clusterings and layouts (see Sec.~\ref{s:appl}), which
  is crucial because both representations are often used together.
\end{itemize}

\begin{acknowledgments}
The author thanks Martin Junghans, Randolf Rotta, and Frank Steinbr\"uckner
for helpful discussions.
\end{acknowledgments}


\begin{thebibliography}{40}
\expandafter\ifx\csname natexlab\endcsname\relax\def\natexlab#1{#1}\fi
\expandafter\ifx\csname bibnamefont\endcsname\relax
  \def\bibnamefont#1{#1}\fi
\expandafter\ifx\csname bibfnamefont\endcsname\relax
  \def\bibfnamefont#1{#1}\fi
\expandafter\ifx\csname citenamefont\endcsname\relax
  \def\citenamefont#1{#1}\fi
\expandafter\ifx\csname url\endcsname\relax
  \def\url#1{\texttt{#1}}\fi
\expandafter\ifx\csname urlprefix\endcsname\relax\def\urlprefix{URL }\fi
\providecommand{\bibinfo}[2]{#2}
\providecommand{\eprint}[2][]{\url{#2}}

\bibitem[{\citenamefont{Simon}(1962)}]{Simon:1962}
\bibinfo{author}{\bibfnamefont{H.~A.} \bibnamefont{Simon}},
  \bibinfo{journal}{Proc.\ Am.\ Philos.\ Soc.}
  \textbf{\bibinfo{volume}{106}}, \bibinfo{pages}{467} (\bibinfo{year}{1962}).

\bibitem[{\citenamefont{Brandes}(2001)}]{Brandes:1999a}
\bibinfo{author}{\bibfnamefont{U.}~\bibnamefont{Brandes}}, in
  \emph{\bibinfo{booktitle}{Drawing Graphs: Methods and Models}}, edited by
  \bibinfo{editor}{\bibfnamefont{M.}~\bibnamefont{Kaufmann}} \bibnamefont{and}
  \bibinfo{editor}{\bibfnamefont{D.}~\bibnamefont{Wagner}}
  (\bibinfo{publisher}{Springer}, \bibinfo{address}{Berlin},
  \bibinfo{year}{2001}), LNCS 2025, pp. \bibinfo{pages}{71--86}.

\bibitem[{\citenamefont{{Di Battista} et~al.}(1999)\citenamefont{{Di Battista},
  Eades, Tamassia, and Tollis}}]{DiBattistaEtAl:1999}
\bibinfo{author}{\bibfnamefont{G.}~\bibnamefont{{Di Battista}}},
  \bibinfo{author}{\bibfnamefont{P.}~\bibnamefont{Eades}},
  \bibinfo{author}{\bibfnamefont{R.}~\bibnamefont{Tamassia}}, \bibnamefont{and}
  \bibinfo{author}{\bibfnamefont{I.~G.} \bibnamefont{Tollis}},
  \emph{\bibinfo{title}{Graph Drawing: Algorithms for the Visualization of
  Graphs}} (\bibinfo{publisher}{Prentice Hall}, \bibinfo{address}{Upper Saddle
  River}, \bibinfo{year}{1999}).

\bibitem[{\citenamefont{Newman}(2004{\natexlab{a}})}]{Newman:EPJ2004}
\bibinfo{author}{\bibfnamefont{M.~E.~J.} \bibnamefont{Newman}},
  \bibinfo{journal}{Eur.\ Phys.\ J.~B}
  \textbf{\bibinfo{volume}{38}}, \bibinfo{pages}{321}
  (\bibinfo{year}{2004}{\natexlab{a}}).

\bibitem[{\citenamefont{Gaertler}(2005)}]{Gaertler:2005}
\bibinfo{author}{\bibfnamefont{M.}~\bibnamefont{Gaertler}}, in
  \emph{\bibinfo{booktitle}{Network Analysis: Methodological Foundations}},
  edited by \bibinfo{editor}{\bibfnamefont{U.}~\bibnamefont{Brandes}}
  \bibnamefont{and} \bibinfo{editor}{\bibfnamefont{T.}~\bibnamefont{Erlebach}}
  (\bibinfo{publisher}{Springer}, \bibinfo{address}{Berlin},
  \bibinfo{year}{2005}), LNCS 3418, pp. \bibinfo{pages}{178--215}.

\bibitem[{\citenamefont{Schaeffer}(2007)}]{Schaeffer:CSR2007}
\bibinfo{author}{\bibfnamefont{S.~E.} \bibnamefont{Schaeffer}},
  \bibinfo{journal}{Comput.\ Sci.\ Rev.} \textbf{\bibinfo{volume}{1}},
  \bibinfo{pages}{27} (\bibinfo{year}{2007}).

\bibitem[{\citenamefont{Fortunato and
  Castellano}(2007)}]{FortunatoCastellano:2007}
\bibinfo{author}{\bibfnamefont{S.}~\bibnamefont{Fortunato}} \bibnamefont{and}
  \bibinfo{author}{\bibfnamefont{C.}~\bibnamefont{Castellano}}
  (\bibinfo{year}{2007}), \bibinfo{note}{preprint arXiv:0712.2716}.

\bibitem[{\citenamefont{Newman and Girvan}(2004)}]{NewmanGirvan:2004}
\bibinfo{author}{\bibfnamefont{M.~E.~J.} \bibnamefont{Newman}}
  \bibnamefont{and} \bibinfo{author}{\bibfnamefont{M.}~\bibnamefont{Girvan}},
  \bibinfo{journal}{Phys.\ Rev.~E} \textbf{\bibinfo{volume}{69}},
  \bibinfo{pages}{026113} (\bibinfo{year}{2004}).

\bibitem[{\citenamefont{Borg and Groenen}(1997)}]{BorgGroenen:1997}
\bibinfo{author}{\bibfnamefont{I.}~\bibnamefont{Borg}} \bibnamefont{and}
  \bibinfo{author}{\bibfnamefont{P.}~\bibnamefont{Groenen}},
  \emph{\bibinfo{title}{Modern Multidimensional Scaling: Theory and
  Applications}} (\bibinfo{publisher}{Springer}, \bibinfo{address}{New
  York}, \bibinfo{year}{1997}).

\bibitem[{\citenamefont{Kamada and Kawai}(1989)}]{KamadaKawai:1989}
\bibinfo{author}{\bibfnamefont{T.}~\bibnamefont{Kamada}} \bibnamefont{and}
  \bibinfo{author}{\bibfnamefont{S.}~\bibnamefont{Kawai}},
  \bibinfo{journal}{Inform.\ Process.\ Lett.}
  \textbf{\bibinfo{volume}{31}}, \bibinfo{pages}{7} (\bibinfo{year}{1989}).

\bibitem[{\citenamefont{Fruchterman and
  Reingold}(1991)}]{FruchtermanReingold:1991}
\bibinfo{author}{\bibfnamefont{T.~M.~J.} \bibnamefont{Fruchterman}}
  \bibnamefont{and} \bibinfo{author}{\bibfnamefont{E.~M.}
  \bibnamefont{Reingold}}, \bibinfo{journal}{Software Pract.\
  Exper.} \textbf{\bibinfo{volume}{21}}, \bibinfo{pages}{1129}
  (\bibinfo{year}{1991}).

\bibitem[{\citenamefont{Davidson and Harel}(1996)}]{DavidsonHarel:1996}
\bibinfo{author}{\bibfnamefont{R.}~\bibnamefont{Davidson}} \bibnamefont{and}
  \bibinfo{author}{\bibfnamefont{D.}~\bibnamefont{Harel}},
  \bibinfo{journal}{ACM Trans.\ Graphics}
  \textbf{\bibinfo{volume}{15}}, \bibinfo{pages}{301} (\bibinfo{year}{1996}).

\bibitem[{\citenamefont{Noack}(2004)}]{Noack:GD2003}
\bibinfo{author}{\bibfnamefont{A.}~\bibnamefont{Noack}}, in
  \emph{\bibinfo{booktitle}{Proceedings of the 11th International Symposium on
  Graph Drawing (GD 2003)}}, edited by
  \bibinfo{editor}{\bibfnamefont{G.}~\bibnamefont{Liotta}}
  (\bibinfo{publisher}{Springer}, \bibinfo{address}{Berlin},
  \bibinfo{year}{2004}), LNCS 2912, pp. \bibinfo{pages}{425--436}.

\bibitem[{\citenamefont{Noack}(2007{\natexlab{a}})}]{Noack:JGAA2007}
\bibinfo{author}{\bibfnamefont{A.}~\bibnamefont{Noack}},
  \bibinfo{journal}{J.\ Graph Algorithms Appl.}
  \textbf{\bibinfo{volume}{11}}, \bibinfo{pages}{453}
  (\bibinfo{year}{2007}{\natexlab{a}}).

\bibitem[{\citenamefont{Noack}(2007{\natexlab{b}})}]{Noack:Diss2007en}
\bibinfo{author}{\bibfnamefont{A.}~\bibnamefont{Noack}}, Ph.D. thesis,
  \bibinfo{school}{Brandenburg University of Technology}
  (\bibinfo{year}{2007}{\natexlab{b}}),
  \urlprefix\url{http://nbn-resolving.de/urn:nbn:de:kobv:co1-opus-4046}.

\bibitem[{\citenamefont{Newman}(2004{\natexlab{b}})}]{Newman:2004}
\bibinfo{author}{\bibfnamefont{M.~E.~J.} \bibnamefont{Newman}},
  \bibinfo{journal}{Phys.\ Rev.~E} \textbf{\bibinfo{volume}{70}},
  \bibinfo{pages}{056131} (\bibinfo{year}{2004}{\natexlab{b}}).

\bibitem[{\citenamefont{Brandes et~al.}(2008)\citenamefont{Brandes, Delling,
  Gaertler, G\"orke, Hoefer, Nikoloski, and Wagner}}]{BrandesEtAl:TKDE2008}
\bibinfo{author}{\bibfnamefont{U.}~\bibnamefont{Brandes}},
  \bibinfo{author}{\bibfnamefont{D.}~\bibnamefont{Delling}},
  \bibinfo{author}{\bibfnamefont{M.}~\bibnamefont{Gaertler}},
  \bibinfo{author}{\bibfnamefont{R.}~\bibnamefont{G\"orke}},
  \bibinfo{author}{\bibfnamefont{M.}~\bibnamefont{Hoefer}},
  \bibinfo{author}{\bibfnamefont{Z.}~\bibnamefont{Nikoloski}},
  \bibnamefont{and} \bibinfo{author}{\bibfnamefont{D.}~\bibnamefont{Wagner}},
  \bibinfo{journal}{IEEE Trans.\ Knowl.\ Data Eng.}
  \textbf{\bibinfo{volume}{20}}, \bibinfo{pages}{172} (\bibinfo{year}{2008}).

\bibitem[{\citenamefont{Hachul and J\"unger}(2007)}]{HachulJunger:JGAA2007}
\bibinfo{author}{\bibfnamefont{S.}~\bibnamefont{Hachul}} \bibnamefont{and}
  \bibinfo{author}{\bibfnamefont{M.}~\bibnamefont{J\"unger}},
  \bibinfo{journal}{J.\ Graph Algorithms Appl.}
  \textbf{\bibinfo{volume}{11}}, \bibinfo{pages}{345} (\bibinfo{year}{2007}).

\bibitem[{\citenamefont{Barnes and Hut}(1986)}]{BarnesHut:1986}
\bibinfo{author}{\bibfnamefont{J.}~\bibnamefont{Barnes}} \bibnamefont{and}
  \bibinfo{author}{\bibfnamefont{P.}~\bibnamefont{Hut}},
  \bibinfo{journal}{Nature} \textbf{\bibinfo{volume}{324}},
  \bibinfo{pages}{446} (\bibinfo{year}{1986}).

\bibitem[{\citenamefont{Danon et~al.}(2005)\citenamefont{Danon,
  D{\'i}az-Guilera, Duch, and Arenas}}]{DanonEtAl:2005}
\bibinfo{author}{\bibfnamefont{L.}~\bibnamefont{Danon}},
  \bibinfo{author}{\bibfnamefont{A.}~\bibnamefont{D{\'i}az-Guilera}},
  \bibinfo{author}{\bibfnamefont{J.}~\bibnamefont{Duch}}, \bibnamefont{and}
  \bibinfo{author}{\bibfnamefont{A.}~\bibnamefont{Arenas}},
  \bibinfo{journal}{J.~Stat.~Mech.},
  \bibinfo{pages}{P09008} (\bibinfo{year}{2005}).


\bibitem[{\citenamefont{Schuetz and
  Caflisch}(2008)}]{SchuetzCaflisch:PRE2008}
\bibinfo{author}{\bibfnamefont{P.}~\bibnamefont{Schuetz}} \bibnamefont{and}
  \bibinfo{author}{\bibfnamefont{A.}~\bibnamefont{Caflisch}},
  \bibinfo{journal}{Phys.\ Rev.~E} \textbf{\bibinfo{volume}{77}},
  \bibinfo{pages}{046112} (\bibinfo{year}{2008}).

\bibitem[{\citenamefont{Rotta}(2008{\natexlab{b}})}]{Rotta:DA2008}
\bibinfo{author}{\bibfnamefont{R.}~\bibnamefont{Rotta}}, Diploma thesis,
  \bibinfo{school}{Brandenburg University of Technology}
  (\bibinfo{year}{2008}{\natexlab{b}}),
  \urlprefix\url{http://www.informatik.tu-cottbus.de/~rrotta/}.

\bibitem[{\citenamefont{Diestel}(2000)}]{Diestel:2000}
\bibinfo{author}{\bibfnamefont{R.}~\bibnamefont{Diestel}},
  \emph{\bibinfo{title}{Graph Theory}} (\bibinfo{publisher}{Springer},
  \bibinfo{address}{New York}, \bibinfo{year}{2000}), \bibinfo{edition}{2nd}
  ed.

\bibitem[{\citenamefont{Reichardt and
  Bornholdt}(2006)}]{ReichardtBornholdt:PR2006}
\bibinfo{author}{\bibfnamefont{J.}~\bibnamefont{Reichardt}} \bibnamefont{and}
  \bibinfo{author}{\bibfnamefont{S.}~\bibnamefont{Bornholdt}},
  \bibinfo{journal}{Phys.\ Rev.~E} \textbf{\bibinfo{volume}{74}},
  \bibinfo{pages}{016110} (\bibinfo{year}{2006}).

\bibitem[{\citenamefont{Fortunato and
  Barth{\'e}lemy}(2007)}]{FortunatoBarthelemy:PNAS2007}
\bibinfo{author}{\bibfnamefont{S.}~\bibnamefont{Fortunato}} \bibnamefont{and}
  \bibinfo{author}{\bibfnamefont{M.}~\bibnamefont{Barth{\'e}lemy}},
  \bibinfo{journal}{Proc.\ Natl.\ Acad.\ Sci.\ U.S.A.}
  \textbf{\bibinfo{volume}{104}}, \bibinfo{pages}{36} (\bibinfo{year}{2007}).

\bibitem[{\citenamefont{Eades}(1984)}]{Eades:1984}
\bibinfo{author}{\bibfnamefont{P.}~\bibnamefont{Eades}},
  \bibinfo{journal}{Congressus Numerantium} \textbf{\bibinfo{volume}{42}},
  \bibinfo{pages}{149} (\bibinfo{year}{1984}).

\bibitem[{\citenamefont{Noack}(2006)}]{Noack:GD2005}
\bibinfo{author}{\bibfnamefont{A.}~\bibnamefont{Noack}}, in
  \emph{\bibinfo{booktitle}{Proceedings of the 13th International Symposium on
  Graph Drawing (GD 2005)}}, edited by
  \bibinfo{editor}{\bibfnamefont{P.}~\bibnamefont{Healy}} \bibnamefont{and}
  \bibinfo{editor}{\bibfnamefont{N.~S.} \bibnamefont{Nikolov}}
  (\bibinfo{publisher}{Springer}, \bibinfo{address}{Berlin},
  \bibinfo{year}{2006}), LNCS 3843, pp. \bibinfo{pages}{309--320}.

\bibitem[{\citenamefont{Shepard}(1980)}]{Shepard:1980}
\bibinfo{author}{\bibfnamefont{R.~N.} \bibnamefont{Shepard}},
  \bibinfo{journal}{Science} \textbf{\bibinfo{volume}{210}},
  \bibinfo{pages}{390} (\bibinfo{year}{1980}).

\bibitem[{\citenamefont{Carroll and Pruzansky}(1980)}]{CarrollPruzansky:1980}
\bibinfo{author}{\bibfnamefont{J.~D.} \bibnamefont{Carroll}} \bibnamefont{and}
  \bibinfo{author}{\bibfnamefont{S.}~\bibnamefont{Pruzansky}}, in
  \emph{\bibinfo{booktitle}{Similarity and Choice: Papers in Honour of Clyde
  Coombs}}, edited by \bibinfo{editor}{\bibfnamefont{E.~D.}
  \bibnamefont{Lantermann}} \bibnamefont{and}
  \bibinfo{editor}{\bibfnamefont{H.}~\bibnamefont{Feger}}
  (\bibinfo{publisher}{Huber}, \bibinfo{address}{Bern}, \bibinfo{year}{1980}),
  pp. \bibinfo{pages}{108--139}.

\bibitem[{\citenamefont{White and Smyth}(2005)}]{WhiteSmyth:2005long}
\bibinfo{author}{\bibfnamefont{S.}~\bibnamefont{White}} \bibnamefont{and}
  \bibinfo{author}{\bibfnamefont{P.}~\bibnamefont{Smyth}}, in
  \emph{\bibinfo{booktitle}{Proceedings of the 5th SIAM International
  Conference on Data Mining (SDM 2005)}} (\bibinfo{publisher}{SIAM},
  \bibinfo{address}{Philadelphia}, \bibinfo{year}{2005}), pp.
  \bibinfo{pages}{274--285}.

\bibitem[{\citenamefont{Newman}(2006{\natexlab{a}})}]{Newman:PR2006}
\bibinfo{author}{\bibfnamefont{M.~E.~J.} \bibnamefont{Newman}},
  \bibinfo{journal}{Phys.\ Rev.~E} \textbf{\bibinfo{volume}{74}},
  \bibinfo{pages}{036104} (\bibinfo{year}{2006}{\natexlab{a}}).

\bibitem[{\citenamefont{Linial et~al.}(1995)\citenamefont{Linial, London, and
  Rabinovich}}]{LinialEtAl:1995}
\bibinfo{author}{\bibfnamefont{N.}~\bibnamefont{Linial}},
  \bibinfo{author}{\bibfnamefont{E.}~\bibnamefont{London}}, \bibnamefont{and}
  \bibinfo{author}{\bibfnamefont{Y.}~\bibnamefont{Rabinovich}},
  \bibinfo{journal}{Combinatorica} \textbf{\bibinfo{volume}{15}},
  \bibinfo{pages}{215} (\bibinfo{year}{1995}).

\bibitem[{\citenamefont{Aumann and Rabani}(1998)}]{AumannRabani:1998}
\bibinfo{author}{\bibfnamefont{Y.}~\bibnamefont{Aumann}} \bibnamefont{and}
  \bibinfo{author}{\bibfnamefont{Y.}~\bibnamefont{Rabani}},
  \bibinfo{journal}{SIAM J.\ Comput.} \textbf{\bibinfo{volume}{27}},
  \bibinfo{pages}{291} (\bibinfo{year}{1998}).

\bibitem[{\citenamefont{Arora et~al.}(2004)\citenamefont{Arora, Rao, and
  Vazirani}}]{AroraEtAl:2004long}
\bibinfo{author}{\bibfnamefont{S.}~\bibnamefont{Arora}},
  \bibinfo{author}{\bibfnamefont{S.}~\bibnamefont{Rao}}, \bibnamefont{and}
  \bibinfo{author}{\bibfnamefont{U.~V.} \bibnamefont{Vazirani}}, in
  \emph{\bibinfo{booktitle}{Proceedings of the 36th ACM Symposium on
  Theory of Computing (STOC 2004)}}, edited by
  \bibinfo{editor}{\bibfnamefont{L.}~\bibnamefont{Babai}}
  (\bibinfo{publisher}{ACM}, \bibinfo{address}{New York},
  \bibinfo{year}{2004}), pp. \bibinfo{pages}{222--231}.

\bibitem[{\citenamefont{Zachary}(1977)}]{Zachary:1977}
\bibinfo{author}{\bibfnamefont{W.~W.} \bibnamefont{Zachary}},
  \bibinfo{journal}{J.~Anthropol.\ Res.}
  \textbf{\bibinfo{volume}{33}}, \bibinfo{pages}{452} (\bibinfo{year}{1977}).

\bibitem[{\citenamefont{Newman}(2006{\natexlab{b}})}]{Newman:PNAS2006}
\bibinfo{author}{\bibfnamefont{M.~E.~J.} \bibnamefont{Newman}},
  \bibinfo{journal}{Proc.\ Natl.\ Acad.\ Sci.\ U.S.A.}
  \textbf{\bibinfo{volume}{103}}, \bibinfo{pages}{8577}
  (\bibinfo{year}{2006}{\natexlab{b}}).

\bibitem[{\citenamefont{Ross and Murphy}(1999)}]{RossMurphy:1999}
\bibinfo{author}{\bibfnamefont{B.~H.} \bibnamefont{Ross}} \bibnamefont{and}
  \bibinfo{author}{\bibfnamefont{G.~L.} \bibnamefont{Murphy}},
  \bibinfo{journal}{Cognitive Psychol.} \textbf{\bibinfo{volume}{38}},
  \bibinfo{pages}{495} (\bibinfo{year}{1999}).

\bibitem[{\citenamefont{Hubert et~al.}(2001)\citenamefont{Hubert, Arabie, and
  Meulman}}]{HubertEtAl:2001}
\bibinfo{author}{\bibfnamefont{L.}~\bibnamefont{Hubert}},
  \bibinfo{author}{\bibfnamefont{P.}~\bibnamefont{Arabie}}, \bibnamefont{and}
  \bibinfo{author}{\bibfnamefont{J.}~\bibnamefont{Meulman}},
  \emph{\bibinfo{title}{Combinatorial Data Analysis: Optimization by Dynamic
  Programming}} (\bibinfo{publisher}{SIAM}, \bibinfo{address}{Philadelphia}, 
  \bibinfo{year}{2001}).

\bibitem[{\citenamefont{Leicht and Newman}(2008)}]{LeichtNewman:PRL2008}
\bibinfo{author}{\bibfnamefont{E.~A.} \bibnamefont{Leicht}} \bibnamefont{and}
  \bibinfo{author}{\bibfnamefont{M.~E.~J.} \bibnamefont{Newman}},
  \bibinfo{journal}{Phys.\ Rev.\ Lett.} \textbf{\bibinfo{volume}{100}},
  \bibinfo{pages}{118703} (\bibinfo{year}{2008}).

\bibitem[{\citenamefont{Barber}(2007)}]{Barber:PRE2007}
\bibinfo{author}{\bibfnamefont{M.~J.} \bibnamefont{Barber}},
  \bibinfo{journal}{Phys.\ Rev.~E} \textbf{\bibinfo{volume}{76}},
  \bibinfo{pages}{066102} (\bibinfo{year}{2007}).

\bibitem[{\citenamefont{Gaertler et~al.}(2007)\citenamefont{Gaertler, G\"orke,
  and Wagner}}]{GaertlerEtAl:AAIM2007}
\bibinfo{author}{\bibfnamefont{M.}~\bibnamefont{Gaertler}},
  \bibinfo{author}{\bibfnamefont{R.}~\bibnamefont{G\"orke}}, \bibnamefont{and}
  \bibinfo{author}{\bibfnamefont{D.}~\bibnamefont{Wagner}}, in
  \emph{\bibinfo{booktitle}{Proceedings of the 3rd International Conference on
  Algorithmic Aspects in Information and Management (AAIM 2007)}}, edited by
  \bibinfo{editor}{\bibfnamefont{M.-Y.} \bibnamefont{Kao}} \bibnamefont{and}
  \bibinfo{editor}{\bibfnamefont{X.-Y.} \bibnamefont{Li}}
  (\bibinfo{publisher}{Springer}, \bibinfo{address}{Berlin},
  \bibinfo{year}{2007}), LNCS 4508, pp. \bibinfo{pages}{11--26}.

\end{thebibliography}

\end{document}